\documentclass{article} 
\usepackage{graphicx}
\usepackage{amscd}
\usepackage{amsmath}
\usepackage{amsfonts}
\usepackage{amssymb}

\begin{document}
\title{Energy distribution in the dyadosphere of a charged black hole}
\author{S.~S.~Xulu \\
 Department Computer Science\\
  University of Zululand\\
  Private Bag X1001\\
  3886 Kwa-Dlangezwa\\
   South Africa}



\maketitle

\begin{abstract}

The event horizon of a charged black hole is, according to
Ruffini\cite{ Ruffini} and Preparata \emph{et
al.}\cite{PreparataEtAl}, surrounded by a special region called
the \emph{dyadosphere} where the electromagnetic field exceeds the
Euler-Heisenberg critical value for electron-positron pair
production. We obtain the energy distribution in the dyadosphere
region for a Reissner-Nordstr\"{o}m black hole.  We find that the
energy-momentum prescriptions of Einstein, Landau-Lifshitz,
Papapetrou, and Weinberg give the same and acceptable energy
distribution.
\end{abstract}



\section{Introduction} \label{sec:intro}

Ruffini\cite{Ruffini}  introduced a new concept of
\emph{dyadosphere} of an electromagnetic black hole to explain
gamma ray bursts. He defined the dyadosphere  as the region just
outside the horizon of a charged black hole whose electromagnetic
field strength is larger than the well-known Heisenbeg-Euler
critical value
\begin{equation}\label{EulerHeisEq}
  \mathcal{E}_{crit}=\frac{m_{e}^{2}c^{3}}{e\hbar}
\end{equation}
for electron-positron pair production ($m_e$ and $e$ respectively
denote mass and charge of an electron.) For a
Reissner-Nordstr\"{o}m space-time, the dyadosphere is described by
the radial interval $r_{+}\leq r \leq r_{ds}$ where the horizon
\begin{equation}
r_{+}=\frac{GM}{c^2}\left(1+\sqrt{1-\frac{q^2}{GM^2}}\right),
\end{equation}
forms the inner radius of the dyadosphere, while its outer radius
is given by
\begin{equation}
r_{ds}=\sqrt{\left( \frac{\hbar}{m_{e}c}\right)\left(
\frac{GM}{c^2}\right)\left( \frac{m_p}{m_e}\right)\left(
\frac{e}{q_p}\right)\left( \frac{Q}{\sqrt{G}M}\right)},
\end{equation}
where $M$ and $Q$ are mass and charge parameters, $m_p=\sqrt{\hbar
c/G }$ and $q_c = \sqrt{\hbar c}$ are respectively the Planck mass
and Planck charge (see in \cite{Ruffini}). Ruffini\cite{Ruffini},
and Preparata \emph{et al.}\cite{PreparataEtAl} have investigated
certain properties of the dyadosphere corresponding to the
Reissner-Nordstr\"{o}m space-time (see also Ruffini \emph{et
al.}\cite{RuffiniEtAl}). They\cite{RuffiniEtAl} found that the
electron-positron pair creation process occur over the entire
dyadosphere, excluding the horizon, where the electromagnetic
field is Heisenberg-Euler overcritical. The total energy of
electron-positron pairs converted from static electric energy and
deposited within the dyadosphere is calculated\cite{RuffiniEtAl}
to be:
\begin{equation}
E_{dya} = \frac{1}{2}\frac{Q^2}{r_+}(1-\frac{r_+}{r_{ds}})\left[1
- \left(\frac{r_+}{r_{ds}}\right)^2\right]. \label{energyDYA}
\end{equation}
Several studies show that, in the presence of  a strong
electromagnetic field,  the velocity of light propagation depends
on vacuum polarization states (Birula and Birula\cite{BIrula},
Adler\cite{Adler}, De Lorenci \emph{et al.}\cite{DeLorenci2000}).
Drummond and Hathrell\cite{DrummHathrell} showed that the effect
of vacuum polarization may lead to superluminal photon
propagation. Daniels and Shore\cite{DanShore} investigated photon
propagation around a charged black hole. Their
results\cite{DanShore} show that the effect of one-loop vacuum
polarization on photon propagation in Reissner-Nordstr\"{o}m
space-time makes superluminal photon propagation possible. Vacuum
polarisation effects thus violate the Principle of Equivalence in
interacting field theories.  It is therefore of interest to
examine further properties of the region where the electromagnetic
field exceeds the Euler-Heissenberg critical limit. In this paper
we investigate the energy distribution in the dyadosphere of a
Reissner-Nordstr\"{o}m space-time using the energy-momentum
complexes of Einstein, Landau-Lifshitz, Papapetrou, and Weinberg.

Einstein's formulation of energy-momentum covariant conservation
laws $\nabla_{a}T^{ab}=0$ ($T^{ab}$ is the energy-momentum tensor
of matter and all nongravitational fields) in the form of Poynting
theorem ($\partial_a \Theta^a_b = 0$, where  $\Theta^a_b$ is known
as the Einstein energy-momentum complex) to include contributions
from gravitational field involved the introduction of a
pseudotensorial quantity. Owing to the fact that
$\Theta_{a}{}_{b}$ is not a true tensor (although covariant under
linear transformations), Levi-Civita, Schr\"{o}dinger, and Bauer
expressed some doubts at the importance of Einstein's local
energy-momentum conservation laws (see in Cattani and De Maria
\cite{CattDeMaria}).  Einstein showed that his energy-momentum
complex provides satisfactory expression for the total energy and
momentum of isolated systems. This was followed by many
prescriptions: e.g. Landau and Lifshitz, Papapetrou, Weinberg, and
many others (see in Refs. \cite{ACV96,BrownYork}). Most of these
prescriptions are coordinate dependent and others are not. The
physical meaning of these was questioned, and the large number of
energy-momentum prescriptions not only fuelled scepticism that
different energy-momentum definitions could give unacceptable
different energy distributions for a given space-time, but also
lead to diverse viewpoints on the possibility of localization of
energy-momentum. In a series of papers,
Cooperstock\cite{CoopHyp,CoopRec1,CoopRec2} hypothesized that in
general relativity energy and momentum are located only to the
regions of nonvanishing energy-momentum tensor and that
consequently the gravitational waves are not carriers of energy
and momentum. Although recent results of Xulu\cite{Xulu6} and
Bringley\cite{Bringley} support this hypothesis, further
investigation of this hypothesis is still required.

The main weaknesses of energy-momentum complexes is that most of
these restrict one to make calculations in ``Cartesian
coordinates''. The alternative concept of quasi-local mass is more
attractive because these are not restricted to the use of any
special coordinate system. There is also a large number of
definitions of quasi-local masses. It has been shown\cite{Bergqv}
that for a given space-time, many quasi-local mass definitions do
not give agreed results. On the other hand, the pioneering
contributions of Virbhadra and co-workers (\cite{ACV96, KSVpapers,
KSVColl})  encouarged numerous researchers (see
\cite{Bringley,Xulu345,YangRadinschi} and  references therein) to
show that for many  space-times several energy-momentum complexes
give the same and acceptable energy-momentum distribution for a
given space-time. Inspired by Virbhadra's result, Chang, Nester
and Chen\cite{ChangEtAl} demonstrated that by associating each of
the energy-momentum complexes of Einstein, Landau and Lifshitz,
M\o ller, Papapetrou, and Weinberg with a legitimate Hamiltonian
boundary term, each of these complexes may be said to be
quasi-local.  Quasi-local energy-momentum are obtainable from a
Hamiltonian. This important paper of Chang,\textit{et
al}\cite{ChangEtAl} dispels the doubts about the physical meaning
of these energy-momentum complexes. Hence, below we use
energy-momentum complexes to obtain energy distribution in the
dyadosphere of a Reissner-Nordstr\"{o}m space-time. In the rest of
this paper we use $G=1,c=1$ units and follow the convention that
Latin indices take values from $0$ to $3$ and Greek indices take
values from $1$ to $3$.

\section{The De Lorenci \emph{et al.} Metric}

De Lorenci, Figueiredo, Fliche and Novello \cite{DeLorenciEtFFN}
calculated the correction for the Reissner-Nordstr\"{o}m metric
from the first contribution of the Euler-Heisenberg Lagrangian and
obtained the following metric
\begin{equation}
ds^{2}= \left( 1-\frac{2M}{r} + \frac{ Q^2}{ r^2} - \frac{ \sigma
Q^4}           {5 r^6}\right)\,dt^{2} - \left( 1-\frac{2M}{r} +
\frac{ Q^2}{ r^2} - \frac{ \sigma Q^4}{5 r^6}
          \right)^{-1}  \,dr^{2}
 - r^{2}\,(d\theta ^{2}+\sin ^{2}\theta \,d\phi ^{2}) \text{.}
\label{Metric}
\end{equation}
By making $\sigma = 0$ we again obtain the Reissner-Nordstr\"{o}m
metric. De Lorenci \emph{et al.}\cite{DeLorenciEtFFN} showed that
the correction term $\frac{ \sigma Q^4}{5 r^6}$ is of the same
order of magnitude as the Reissner-Nordstr\"{o}m charge term
$\frac{ Q^2}{2 r^2}$.

In order to compute the energy distribution using the
energy-momentum complexes of Einstein, Landau-Lifshitz,
Papapetrou, and Weinberg we are restricted to the use of
``Cartesian coordinates''. Therefore we can express the above
metric (\ref{Metric}) in ${T,x,y,z}$ coordinates. The coordinates
transformation is:
\begin{eqnarray}
T &=& t+r - \int \left( 1-\frac{2M}{r} + \frac{ Q^2}{ r^2} -
\frac{ \sigma Q^4}{5 r^6}
          \right)^{-1}  \,dr , \nonumber\\
x\ &=& \ r\ \sin\theta\ \cos\phi, \nonumber\\
y\ &=& \ r\ \sin\theta \ \sin\phi, \nonumber\\
z\ &=& \ r\ \cos\theta \text{.} \label{}
\end{eqnarray}
Thus one has the metric in ${T,x,y,z}$ coordinates:
\begin{eqnarray}
ds^2 &=& dT^2 - dx^2 -  dy^2 - dz^2 - \nonumber\\
    && \left(\frac{2M}{r} - \frac{ Q^2}{ r^2} + \frac{ \sigma Q^4}
    {5 r^6}\right) \times \left[dT - \frac{x dx + y dy +z dz}{r}\right]^2.
\label{MetricCart}
\end{eqnarray}

\section{Einstein energy-momentum complex}

The  Einstein energy-momentum complex is given as
\begin{equation}
\Theta_i{}^{k} = \frac{1}{16 \pi} H^{\ kl}_{i \ \ ,l} ,
\label{EinsteinComplex}
\end{equation}
where
\begin{equation}
H_i^{\ kl}  =  - H_i^{\ lk}\ =\  \frac{g_{in}}{\sqrt{-g}}
         \left[-g \left( g^{kn} g^{lm} - g^{ln} g^{km}\right)\right]_{,m} \ .
\end{equation}
$\Theta_0^{\ 0}$ and $\Theta_{\alpha}^{\ 0}$ denote for the energy
and momentum density components, respectively.
(Virbhadra\cite{KSV99} mentioned that though the  energy-momentum
complex found by Tolman differs in form from the Einstein
energy-momentum complex, both are equivalent in import.) The
energy-momentum components are expressed by
\begin{equation}
P_i  =  \int \int \int \Theta_i^{\ 0} dx^1  dx^2 dx^3 .
\end{equation}
Further Gauss's theorem  furnishes
\begin{equation}
P_i  = \frac{1}{16 \pi} \ \int\int\ H_i^{\ 0 \alpha} \
\mu_{\alpha}\ dS \text{,}\label{GaussE}
\end{equation}
where $\mu_{\alpha}$ is the outward unit normal vector over the
infinitesimal surface element $dS$.  $P_\alpha$ give momentum
components $P_1, P_2, P_3$ and $P_{0}$ gives the energy.

The only required components of $H_i^{\ j k}$ in the calculation
of energy are the following:
\begin{eqnarray}
H_0^{\ 0 1} &=& \gamma x \text{,} \nonumber \\
H_0^{\ 0 2} &=& \gamma y \text{,} \nonumber \\
H_0^{\ 0 3} &=& \gamma z \text{,} \label{Hcomponents}
\end{eqnarray}
where
\begin{equation}
\gamma = \frac{4M}{r^{3}} - \frac{2 Q^{2}}{r^{4}} +
\frac{2Q^{4}\sigma}{5r^{8}} .\label{gamma}
\end{equation}
For a surface given by parametric equations $x=r\sin \theta \cos
\phi ,$ $\ y=r\sin \theta \sin \phi ,$ $\ z=r\cos \theta $ (where
$r$ is constant) one has $\mu _{\beta }=\{x/r,$ $y/r,$ $z/r\}$ and
$dS=r^{2}sin\theta d\theta d\phi $. Now using
($\ref{Hcomponents}$) in ($\ref{GaussE}$) over a surface
$r=const.$, we obtain:
\begin{equation}
E_{\rm Einst} =
         M  -  \frac{Q^2}{2 r} +  \frac{\sigma Q^4}{10 r^5}
\label{EEinst}
\end{equation}
In the next Section we obtain the energy distribution for the same
metric in Landau and Lifshitz formulation.

\section{Energy distribution in Landau and Lifshitz Formulation}

The symmetric energy-momentum complex of Landau and
Lifshitz\cite{LL} may be written as
\begin{equation}
L^{ij}=\frac{1}{16\pi }\ell_{\quad ,kl}^{ikjl}  \label{LLcomplex}
\end{equation}
where
\begin{equation}
\ell^{ikjl}=-g(g^{ij}g^{kl}-g^{il}g^{kj}).  \label{Sijkl}
\end{equation}
$L^{00}$ is the energy density and $L^{0\alpha }$ are the momentum
(energy current) density components. $ \ell ^{mjnk} $ has
symmetries of the Riemann curvature tensor. The expression
\begin{equation}
P^i = \int\int\int L^{i0} dx^1 dx^2 dx^3
\end{equation}
gives the energy $P^0$ and momentum $P^\alpha$ components. Thus
after applying the Gauss theorem, the energy expression is given
by
\begin{equation}
E_{LL}=\frac{1}{16\pi }\int \int \ell_{\quad ,\alpha }^{0\alpha
0\beta }\ \mu _{\beta }\ dS  ,\label{LLGauss}
\end{equation}
where $\mu _{\beta }$ is the outward unit normal vector over an
infinitesimal surface element $dS$. In order to calculate the
energy component for the De Lorenci \emph{et al.} metric expressed
by the line element $(\ref{MetricCart})$, we need the following
non-zero components of $\ell ^{ikjl}$
\begin{eqnarray}
\ell^{0101} &=& -1 +\left( -1 + \frac{x^{2}}{r^{2}} \right)\gamma
\text{,}
\nonumber \\
\ell^{0102} &=& \frac{x y \gamma}{r^{2}} \text{,} \nonumber
\\
\ell^{0103} &=& \frac{x z \gamma}{r^{2}} \text{,} \nonumber
\\
\ell^{0202} &=& -1 +\left( -1 + \frac{y^{2}}{r^{2}} \right)\gamma
\text{,} \nonumber
\\
\ell^{0203} &=& \frac{y z \gamma}{r^{2}} \text{,} \nonumber
\\
\ell^{0303} &=& -1 +\left( -1 + \frac{z^{2}}{r^{2}} \right)\gamma
\text{,}  \label{Scomponents}
\end{eqnarray}
Equation (\ref{LLcomplex}) with eqs. (\ref{Sijkl}) and
(\ref{Scomponents}) gives the  energy density component:
\begin{equation}
L^{00} = \frac{1}{8\pi}\left[ \frac{Q^2}{r^4} - \frac{Q^4
\sigma}{r^8}\right].
\end{equation}
Using equations $(\ref{Scomponents})$ in $(\ref{LLGauss})$ over a
surface $r=const.$, we obtain:
\begin{equation}
E_{\rm LL} =
         M  -  \frac{Q^2}{2 r} +  \frac{\sigma Q^4}{10 r^5}
\label{ELL}
\end{equation}
Thus we find the same energy distribution we obtained in the last
section.  In the next Section we obtain the energy distribution
for the same metric in Papapetrou formulation.


\section{Energy distribution in Papapetrou formulation}

The Papapetrou energy-momentum complex\cite{Papap}:
\begin{equation}
\Omega ^{ij}=\frac{1}{16\pi }{\cal {N}}_{\quad ,kl}^{ijkl}
\label{Papcomplex}
\end{equation}
where
\begin{equation}
{\cal {N}}^{ijkl}=\sqrt{-g}\left( g^{ij}\eta ^{kl}-g^{ik}\eta
^{jl}+g^{kl}\eta ^{ij}-g^{jl}\eta ^{ik}\right)  \label{Nijkl}
\end{equation}
is also symmetric in its indices. $\Omega^{i0}$  are the energy
and momentum density components. Energy  and momentum components
$P^i$ are given by
\begin{equation} P^i =
\int\int\int \Omega ^{i0} dx^1 dx^2 dx^3.
\end{equation}
Applying the Gauss theorem, the energy $E$ for a stationary metric
is given by the expression
\begin{equation}
E_{P}=\frac{1}{16\pi }\int \int {\cal N}_{\qquad ,\beta
}^{00\alpha \beta }\ \ \mu _{\alpha }dS.  \label{PapGauss}
\end{equation}
To find the energy component of the line element
$(\ref{MetricCart})$, we require the following non-zero components
of $\ {\cal N}^{ijkl}$ :
\begin{eqnarray}
{\cal N}^{0011} &=& -1 - \frac{\gamma}{r} + \frac{\gamma
x^{2}}{r^{3}} \text{,}
\nonumber \\
{\cal N}^{0012} &=& \frac{\gamma x y}{r^{3}} \text{,} \nonumber
\\
{\cal N}^{0013} &=& \frac{\gamma x z}{r^{3}} \text{,} \nonumber
\\
{\cal N}^{0022} &=& -1 - \frac{\gamma}{r} + \frac{\gamma
y^{2}}{r^{3}} \text{,}
\nonumber \\
{\cal N}^{0023} &=& \frac{\gamma y z}{r^{3}} \text{,} \nonumber
\\
{\cal N}^{0033} &=& -1 - \frac{\gamma}{r} + \frac{\gamma
z^{2}}{r^{3}} \text{,}  \label{Ncomponents}
\end{eqnarray}
Using the above results in (\ref{Papcomplex}) and (\ref{Nijkl}),
we obtain
\begin{equation}
\Omega^{00} = \frac{1}{8\pi}\left[\frac{Q^2}{r^4} - \frac{Q^4
\sigma}{r^8}\right]
\end{equation}
Thus we find the same energy density as we obtained in the last
section. We now use Eq. $(\ref{Ncomponents})$ in
$(\ref{PapGauss})$ over a 2-surface (as in the last Section) and
obtain
\begin{equation}
E_{\rm Pap} =
         M  -  \frac{Q^2}{2r} +  \frac{\sigma Q^4}{10 r^5},
\label{EPap}
\end{equation}
which is the same as obtained in the previous section. In the next
Section we obtain the energy distribution for the same metric in
Weinberg formulation.


\section{Energy distribution in Weinberg formulation}

The symmetric energy-momentum complex of Weinberg \cite{Weinberg}
is :
\begin{equation}
W^{ik}=\frac{1}{16\pi }{\triangle}_{\quad ,l}^{ikl}
\label{Wcomplex}
\end{equation}
where
\begin{equation}
{\triangle }^{ikl}=\frac{%
\partial h_{a}^{a}}{\partial x_{i}}\eta ^{lk}-\frac{\partial h_{a}^{a}}{\partial x_{l}}\eta
^{ik}+
\frac{\partial h^{al}}{%
\partial x^{a}}\eta ^{ik}-\frac{\partial h^{ai}}{\partial x^{a}}\eta ^{lk}+\frac{\partial
h^{ik}}{\partial x_{l}}-%
\frac{\partial h^{lk}}{\partial x_{i}} \label{triangle}
\end{equation}
and
\begin{equation}
h_{ij}=g_{ij}-\eta _{ij}\text{.}  \label{hij}
\end{equation}
$\eta _{ij}$ \ is the Minkowski metric. The expression
\begin{equation} P^i =
\int\int\int W^{i0} dx^1 dx^2 dx^3.
\end{equation}
gives the energy $P^0$ and momentum $P^\alpha$ components. Once
more Gauss's theorem furnishes the following expression for the
energy $E$ of a stationary metric:
\begin{equation}
E_{W}=\frac{1}{16\pi }\int \int \triangle ^{\alpha 0k}\mu _{\alpha
} dS.  \label{WeinGauss}
\end{equation}
To find the energy component of the line element
$(\ref{MetricCart})$, we require the following non-zero components
of $\ {\triangle}^{ijk}$:
\begin{eqnarray}
\triangle^{100} &=& \gamma x \text{,}\nonumber \\
\triangle^{200} &=& \gamma y \text{,}\nonumber \\
\triangle^{300} &=& \gamma z \text{,}\label{Wcomponents}
\end{eqnarray}
where $\gamma$ is given by Eq. (\ref{gamma}). To find the energy
density component $W^{00}$, we use Eq. (\ref{Wcomplex}) with
(\ref{triangle}) and (\ref{Wcomponents}) and get
\begin{equation}
W^{00} = \frac{1}{8\pi}\left[ \frac{Q^2}{r^4} - \frac{Q^4
\sigma}{r^8}\right],
\end{equation}
which agrees with energy density components of the Landau-Lifshitz
and the Papapetrou energy-momentum complexes. Now using
($\ref{Wcomponents}$) in ($\ref{WeinGauss}$) we obtain
\begin{equation}
E_{\rm W} =
         M  -  \frac{Q^2}{2 r} +  \frac{\sigma Q^4}{10 r^5} .
\label{EWein}
\end{equation}
which is also the same as in the above sections.


\section{\protect\bigskip Conclusion}

Misner \textit{et al} \cite{MTW} argued that to look for a local
energy-momentum is looking for the right answer to the wrong
question. They further argued that energy is only localizable for
spherical systems. Cooperstock and Sarracino \cite{CoopSar}
countered this point of view, arguing that if energy is
localizable in spherical systems then it is localizable in any
space-times. Bondi\cite{Bondi}  pleaded that a nonlocalizable form
of energy is not admissible in general relativity. The viewpoints
of Misner \textit{et al} discouraged further study of energy
localization and on the other hand an alternative concept of
energy, the so-called quasi-local energy, was developed. To date,
a large number of definitions of quasi-local mass have been
proposed. The uses of quasi-local masses to obtain energy in a
curved space-time are not limited to a particular coordinates
system whereas many energy-momentum complexes are restricted to
the use of ``Cartesian coordinates.'' Penrose\cite {Penrose}
emphasized that quasi-local masses are conceptually very
important. Nevertheless, the present quasi-local mass definitions
still have inadequacies. For instance, Bergqvist\cite{Bergqv}
studied seven quasi-local mass definitions and  concluded that no
two of these definitions give agreed results for the
Reissner-Nordstr\o m and Kerr space-times. The shortcomings of the
seminal quasi-local mass definition of Penrose in handling the
Kerr metric are discussed in Bernstein and Tod\cite {BT}, and in
Virbhadra\cite{KSV99}. On the contrary, the remarkable work of
Virbhadra, and some others, and recent results of Chang, Nester
and Chen have revived the interest in various energy-momentum
complexes. Recently Virbhadra stressed that although the
energy-momentum complexes of Einstein, Landau-Lifshitz,
Papapetrou, and Weinberg are nontensorial (under general
coordinate transformations), these do not violate the principle of
covariance as the equations describing the conservation laws with
these objects are true in any coordinates systems.

In this paper we obtained the energy distribution in De Lorenci
\emph{et al.} spacetime using the energy-momentum complexes of
Einstein, Landau-Lifshitz, Papapetrou, and Weinberg. All four
prescriptions give the same distribution of energy ($E_{Einst} =
E_{LL} = E_{Pap} = E_W$) given as:
\begin{equation}
E =
         M  -  \frac{Q^2}{2 r} +  \frac{\sigma Q^4}{10 r^5}
         \label{energy} \text{.}
\end{equation}
It is obvious that in the dyadospheric region (where $r$ is small)
the last term plays a very important role. As expected, $\sigma =
0$ gives the energy distribution for the Reissner-Nordstr\"{o}m
metric.

\textbf{Acknowledgments} I am   grateful to  Prof.  K. S.
       Virbhadra for  some discussions on dyadosphere.



\bibliographystyle{plain}

\end{document}